\documentclass[dvipdfmx, usenatbib]{pasj01}
\usepackage{url}
\usepackage{comment}
\usepackage{graphicx}
\begin{document}
\SetRunningHead{Photo-$z$ WG}{Photo-$z$ for HSC-SSP}

\title{Photometric Redshifts for the Hyper Suprime-Cam Subaru Strategic Program Data Release 2}

\author{
  Atsushi J. Nishizawa\altaffilmark{1},
  Bau-Ching Hsieh\altaffilmark{2},
  Masayuki Tanaka\altaffilmark{3},
  Tadafumi Takata\altaffilmark{3}
}
\altaffiltext{1}{Institute for Advanced Research, Nagoya University Furocho, Chikusa-ku, Nagoya, 464-8602 Japan}
\altaffiltext{2}{Academia Sinica Institute of Astronomy and Astrophysics, 11F of AS/NTU Astronomy-Mathematics Building, No.1, Sec. 4, Roosevelt Rd, Taipei 10617, Taiwan}
\altaffiltext{3}{National Astronomical Observatory of Japan, 2-21-1 Osawa, Mitaka, Tokyo 181-8588, Japan}

\email{atsushi.nishizawa@iar.nagoya-u.ac.jp}


\maketitle
\newcommand{\commentblue}[1]{\textbf{\color{blue}{#1}}}
\newcommand{\commentred}[1]{\textbf{\color{red}{#1}}}
\newcommand{\bs}[1]{ {\boldsymbol{#1}}}
\newcommand{\photoz}{photo-$z$}
\newcommand{\specz}{spec-$z$}

\begin{abstract}
We present a description of the second data release for the photometric redshift (photo-$z$) of the Subaru Strategic Program for the Hyper-Suprime Cam survey.
Our photo-$z$ products for the entire area in the Data Release 2 are publicly available, and both our point estimate catalog products and full PDFs can be retrieved from the data release site, 
\url{https://hsc-release.mtk.nao.ac.jp/}.
\end{abstract}

\section{Introduction}
\label{sec:introduction}
We present, in this paper, the descriptions of the second public data release of photometric redshift (photo-z) of the HSC survey. The secound data release of the HSC photo-z has been published in January 31st, 2020 which is based on the photometric data collected from 2014 February to August 2018.
To date, the HSC photoz has been widely utilized in various sciences; 
cluster of galaxies \citep{2017ApJ...851..139L, 2018PASJ...70S..22M, 2018PASJ...70S..27M, 2018PASJ...70S..23J, 2018PASJ...70S..12T, 2018PASJ...70S..28M, 2018ApJ...866..140Y, 2019ApJ...875...63M, 2019arXiv190508991K, 2019ApJ...876...40P, 2019PASJ...71...79O, 2019arXiv190902042C, 2019arXiv190910524U, 2019PASJ...71..107M, 2019arXiv191109236O, 2019PASJ...71..112H, 2020MNRAS.tmp..112O}, 
cosmoc weak and strong lensing analysis \citep{2018PASJ...70S..26O, 2018PhRvD..97l3015S, 2018ApJ...867..107W, 2018MNRAS.481..164S, 2019A&A...622A..30S, 2019PASJ...71...43H, 2019arXiv190606041H, 2019MNRAS.486.4365S, 2019ApJ...882...62N, 2019arXiv191102195L, 2019MNRAS.490.5658S}, and 
galaxy and quasar studies \citep{2018PASJ...70S..31O, 2018PASJ...70S..21K, 2018PASJ...70S..33H, 2018ApJ...866..140Y, 2019A&A...622A..30S, 2019ApJ...876..132N, 2019ApJ...876...40P, 2019PASJ...71...60W, 2019ApJS..243...15T, 2019MNRAS.489L..12K, 2019PASJ...71..111I, 2020MNRAS.492.3685H, 2020MNRAS.491.1408M, 2020MNRAS.491.5911O, 2020MNRAS.492.2528L}.

In the secound DR of the HSC photo-z, we include more than twice objects compared to the first DR in wider area in HSC Wide fields and deeper photometry in Deep/UltraDeep fields.
This paper is only include a minor update from the HSC photo-z DR1 \citep{tanaka18} (refered as PDR1 paper), so that readers should refer to the first DR paper as well when using the HSC photo-z data.
The paper is organized as follows. In section \ref{sec:training_validation_and_test_samples}, we describe the construction of spectroscopic redshift sample which is used to calibrate our photo-z. In section \ref{sec:method}, we revisit the method to measure the photo-z for each code. In section \ref{sec:performance}, we describe our metrics to assess the quality of our photo-z, both for point estimates and PDFs. Section \ref{sec:products} gives our data products.

\section{Training, Validation, and Test Samples}
\label{sec:training_validation_and_test_samples}

\subsection{Construction of the training sample}
\label{ssec:construction_of_the_training_sample}

We first collect spectroscopic redshifts from the literature:
zCOSMOS DR3 \citep{lilly09},
UDSz \citep{bradshaw13,mclure13},
3D-HST \citep{skelton14,momcheva16},
FMOS-COSMOS \citep{silverman15},
VVDS \citep{lefevre13},
VIPERS PDR1 \citep{garilli14},
SDSS DR14 \citep{alam15},
GAMA DR2 \citep{liske15},
WiggleZ DR1 \citep{drinkwater10},
DEEP2/3 DR4 \citep{davis03,cooper11,cooper12,newman13},
PRIMUS DR1 \citep{coil11,cool13} and
VANDELS DR2 \citep{pentericci18}.
Those catalogs provide a quality control parameter to evaluate the quality of spec-z. As in the PDR1, we construct the homogenized bit flag and it is summarized below for the catalogs which are newly available in this data release. The following criteria should be satisfied before we match the catalog to the HSC photometric catalog.\\

\noindent
\textbf{Public \specz\ data:}\\
\begin{enumerate}
	\item $0.01 < z < 9$ (no stars, quasars, or failures)
	\item $\sigma_z < 0.005(1+z)$ (error cut)
	\item SDSS/BOSS DR14: $\textrm{\texttt{zWarning}} = 0$ (no apparent issues)
	\item DEEP2/3 DR4: $\textrm{\texttt{qFlag}} = 4$ ($>99.5\%$ confidence)
	\item VANDELS DR2: $\textrm{\texttt{zflg}}>2$ ($>95\%$ confident)
	\item For other spec-z catalogs, we apply $\textrm{\texttt{flag\_homogeneous}==True}$ taken from PDR1 \specz catalog.
\end{enumerate}

\noindent
\textbf{Private spec-z data:}\\
In addition to the public \specz catalogs, we combine private catalogs of COSMOS \specz's (Mara Salvato, private communication) and C3R2 \specz's (Dan Masters, private communication, \cite{masters17,masters19}) exclusively used for our photo-z calibration (these catalogs are not included in the PDR2). 

\noindent
\textbf{Multiband \photoz\ data:}\\
In order to cover the deep photometric data of HSC, we also combine COSMOS2015 multiband \photoz \citep{laigle16}.

Those \specz\ catalogs are matched to the HSC UltraDeep/Deep photometric catalog and subsequently to the HSC Wide catalog which meet the following criteria, 
\begin{enumerate}
    \item \texttt{isprimary} is \texttt{True} (no duplicates)
    \item \texttt{[grizy]\_cmodel\_flux} $>0$
    \item \texttt{[grizy]\_cmodel\_fluxsigma} $>0$
\end{enumerate}
The matching radius is taken to 1 arcsec.

\begin{figure}
  \begin{center}
    \includegraphics[width=\linewidth]{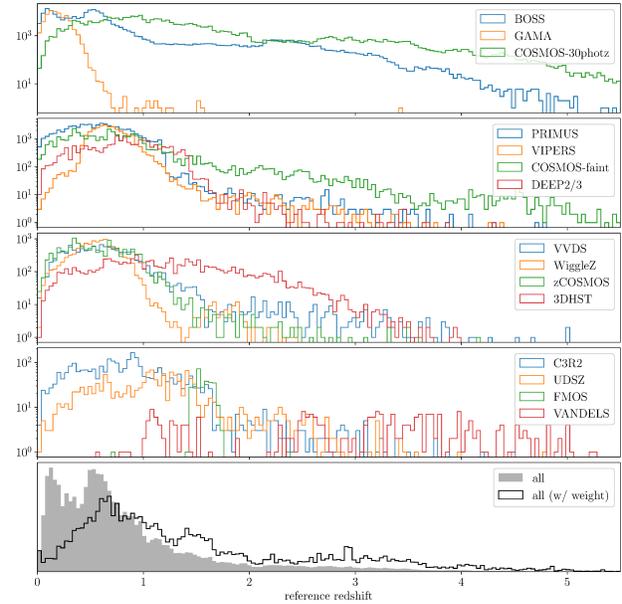}
  \caption{
      Redshift distributions for the training sample. 
  }
 \label{fig:hsc_pz_3}   
  \end{center}  
\end{figure}

The following quantities are then selected and/or computed:
\begin{enumerate}
\item \textbf{Identifiers}: \texttt{ID}, (\texttt{ra,dec}), and (\texttt{tract,patch}) coordinates, but these are confidential during the calibration.
\item \textbf{Fluxes}: \texttt{PSF} fluxes, \texttt{cmodel} fluxes, \texttt{cmodel\_exp} fluxes,
\texttt{cmodel\_dev} fluxes, and PSF-matched aperture fluxes
with target 1.3 arcsec PSF and 1.5 arcsec apertures, together with the Galactic attenuation estimates (\texttt{a\_[grizy]}).
\item \textbf{Shapes}: \texttt{sdss\_shape} parameters.
\item \textbf{Redshift}: redshift, 1$\sigma$ error when available, parent survey (SDSS, etc.).
\item \textbf{Depth}: flag for UltraDeep/Deep, and Wide photometry.
\item \textbf{Emulated errors}: emulated wide-depth photometric errors for the objects which do not match to the Wide layer data. For each UltraDeep/Deep objects, we assign the mean of the photometric errors over wide layer objects located in the nearest neighbors in the magnitude and color hyperspace. 
\item \textbf{Weights}: We compute the specific weights such that the magnitude and magnitude error distribution of the training sample after applying the weights matches to the one for the entire wide layer photometric target sample \citep{lima08}. The weights are thus used to optimize the photo-z to maximize the performance of the target photometric sample, and then used to predict the expected performance of the \photoz\ for the target sample.
\item \textbf{cross-validation ID}: We divide the training sample into 5 subset to conduct a cross validation.
\end{enumerate}
Since we compile spectroscopic, grizm/prism redshifts and high precision photometric redshift as a calibration sample, we refer to those sample as reference redshift $z_{\rm ref}$ hereafter.
Figure \ref{fig:hsc_pz_3} shows the redshift distributions for the reference redshift sample matched to the HSC photometric sample. In the bottom panel, we also show the weighted distribution projected to the entire HSC Wide target sample.

\subsection{Training, validation and test procedures}
\label{ssec:training_validation_proc}
We first divide the calibration sample into $k=5$-fold stratified cross validation set. The sample is equally divided so that the redshift distribution coincide with each other. The 5th sample is reserved for the test sample and is never used for calibration or optimization process.
The rest of the 4 sub-samples are used for training and validation. 
Mizuki uses the 1-hold out validation, i.e. $k=1-3$ are used for training and $k=4$ is for the validation, while DEmP uses $k=1-4$ for both training and leave-one-out validation.
After completing the training, we apply the \photoz\,  code to the test sample. 
Unlike the PDR1 \photoz, we directly evaluate the \photoz\, accuraccy using the $k=5$ validation set, which has not been used for optimizing the code. Also, we do not apply the weight to project the accuracy to the entire photometric target sample.

\section{Methods}
\label{sec:method}
In the PDR2, photo-z catalogs based on one template fitting code \texttt{Mizuki} and one empirical-method-based code \texttt{DEmP} are available. 
Here, we briefly revisit the each \photoz\, measurement method.

\subsection{DEmP}
\label{ssec:demp}

The Direct Empirical Photometric code (\texttt{DEmP}) is an empirical
quadratic polynomial photometric redshift fitting code.
Details are described in \citet{hsieh14} and \citet{tanaka18}.
In PDR2, the photo-$z$, stellar mass, and SFR are provided, and they are computed from afterburner photometry independently.
The photo-$z$, as well as stellar mass and SFR, for each object is calculated using the 40 nearest neighbors 
in a nine-dimensional parameter space (5 magnitude axes and 4 color axes) with a linear function. 
There is no number of filters cut, i.e., the output products are provided for objects with even one-band-only photometry.
The PDF of photo-$z$ for each galaxy is generated using Monte Carlo
technique and the bootstraping method. We use Monte Carlo technique to generate 500 data sets based on
the photometry and uncertainties of the input galaxies to account for the effects due to photometric
uncertainties. We then bootstrap the training set for each input galaxy 500 times for each of
the Monte Carlo generated data set, to estimate the sampling effect in the training set.

\subsection{Mizuki}
\label{ssec:mizuki}

We use a template fitting-code {\sc mizuki}.  Details are described in \citet{tanaka15} and
\citet{tanaka18}, but a brief outline is given here.  The code uses a set of
templates generated with the \citet{bruzual03} stellar population synthesis code assuming
a \citet{chabrier03} IMF and \citet{calzetti00} dust attenuation curve.  Emission
lines are added to the templates assuming solar metallicity \citep{inoue11}.  Template
error functions are incorporated in order to account for systematic offsets and uncertainties
in the templates.  We apply a set of Bayesian priors on the physical properties and let
the priors depend on redshift to keep the template parameters within realistic ranges to
reduce the degeneracy in the multi-color space and also to let templates evolve with redshift
in an observationally motivated way.

An update since PDR1 is that the PDF recalibration is applied following \citet{bordoloi10},
which delivers a small improvement in the overall performance.  We use the CModel photometry,
but as described in \citet{aihara19}, the photometry is occasionally over-estimated due to
severe object blending (a side effect of well-preserved wings of bright objects).
Photo-$z$'s from \texttt{mizuki} suffer from the photometry problem.

\section{Performance Evaluation}
\label{sec:performance}

\subsection{Metrics to characterize photo-$z$}
\label{ssec:metric}


In order for clarity, we explicitly define the quantities to evaluate the performance of the point estimate of the photo-z, which in literature, is often defined in the different manner. We note that unlike in the PDR1, we do not apply a successive three sigma clipping to calculate the following statistics.

\begin{itemize}
    \item {\textbf{Bias $b$}:}\\
    Since the photo-z is not a perfect measure of the redshift of galaxy and often systematically deviates from the spectroscopic redshift. To quantify the difference between photo-z and spec-z, we first define the difference between them as $\Delta z = (z_{\rm photo} - z_{\rm spec})/(1+z_{\rm spec})$. The bias can be defined as the shift averaged over entire sample or some subset of the objects.
    We define here the conventional bais as $b_{\rm conv}\equiv {\rm M}(\Delta z)$, where M stands for the median. Although the median is even stable against the outliers, the better measure of bias can be given by \textit{bi-weighted} mean \citep{beers90},
    \begin{eqnarray}
        &&b_{\rm bw} \equiv M + \frac{\sum_{|u_i|<1}(\Delta z_i - M)(1-u_i^2)^2 }%
        {\sum_{|u_i|<1}(1-u_i^2)^2} \\
        &&u_i = \frac{(\Delta z_i - M)}{c {\rm MAD}},
    \end{eqnarray}
    where MAD is a median absolute deviation, ${\rm MAD}\equiv M(|\Delta z_i-M|)$, and parameter $c$ is tuning parameter and we set it as $c=6.0$. The above formula can be iteratively applyed by replacing $M$ with $b_{\rm bw}$. Convergence of this iteration is quite fast and we iterate till the incremental difference in $b_{\rm bw}$ reaches $0.1\%$.
    
    \item{\textbf{Scatter $\sigma$}:}\\
    Not only the systematic bias, how much the photo-z is scattered around the true spec-z value is also important to explore. It is often in the literature, defined as $\sigma_{\rm conv}={\rm MAD}/0.6745$. Here we also introduce a biweight deviation defined as 
    \begin{eqnarray}
        \sigma_{\rm bw} 
        = 
        \frac{\sqrt{N\sum_{|u_i|<1}(\Delta z_i-M)^2(1-u_i^2)^4}}
        {\left|\sum_{|u_i|<1}(1-u_i^2)(1-5u_i^2)\right|}.
    \end{eqnarray}
    Again, we iterate the measurement simultaneously with $b_{\rm bw}$ by replacing $M$ with $b_{\rm bw}$. Here we set $c=9.0$.
    
    \item{\textbf{outlier rate $\eta$}:}\\
    The outlier is conventionally defined as, $|\Delta z_i| > 0.15$ and thus the outlier rate can be
    \begin{equation}
        \eta_{\rm conv}
        =
        \frac{N(|\Delta z_i|>0.15)}{N_{\rm tot}}.
    \end{equation}
    Here we introduce a biweight based definition of the outlier rate such that 
    \begin{equation}
        \eta
        =
        \frac{N(|\Delta z_i-b_{\rm bw}|>2\sigma_{\rm bw})}{N_{\rm tot}}.
    \end{equation}
    
    \item{\textbf{Loss $L$}:}\\
    Although above statistics are useful to quantify the accuracy of the photo-z, it is more convenient to introduce loss function as it has been introduced in \citet{tanaka18},
    \begin{equation}
        L(\Delta z) = 1 - \frac{1}{1+(\Delta z/\gamma)^2}
    \end{equation}
    This loss function can incorporate bias, scatter and outlier rate at the same time and can evaluate the photo-z accuracy with single quantity.
\end{itemize}

\begin{figure*}
    \centering
    \includegraphics[width=\linewidth]{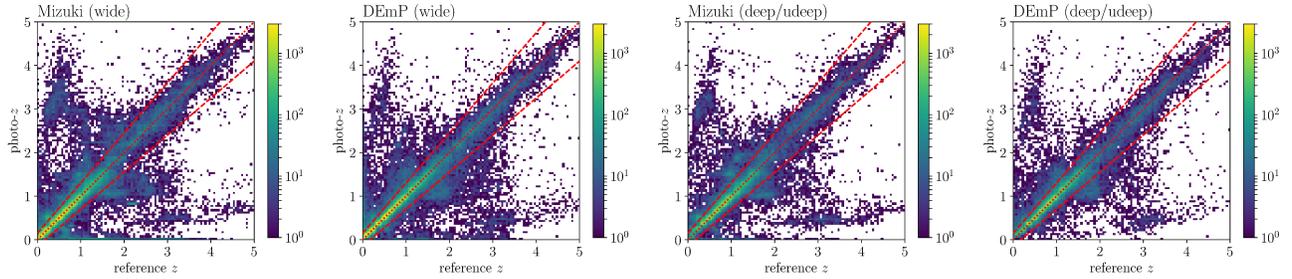}
    \caption{Scatter plot of HSC \photoz\, versus reference redshift. Four different panels show different \photoz\, codes and different depths. For reference, red lines indicate the boundaries of the conventional outlier.}
    \label{fig:scatterPlot}
\end{figure*}
\begin{figure*}
  \begin{center}
    \includegraphics[width=\linewidth]{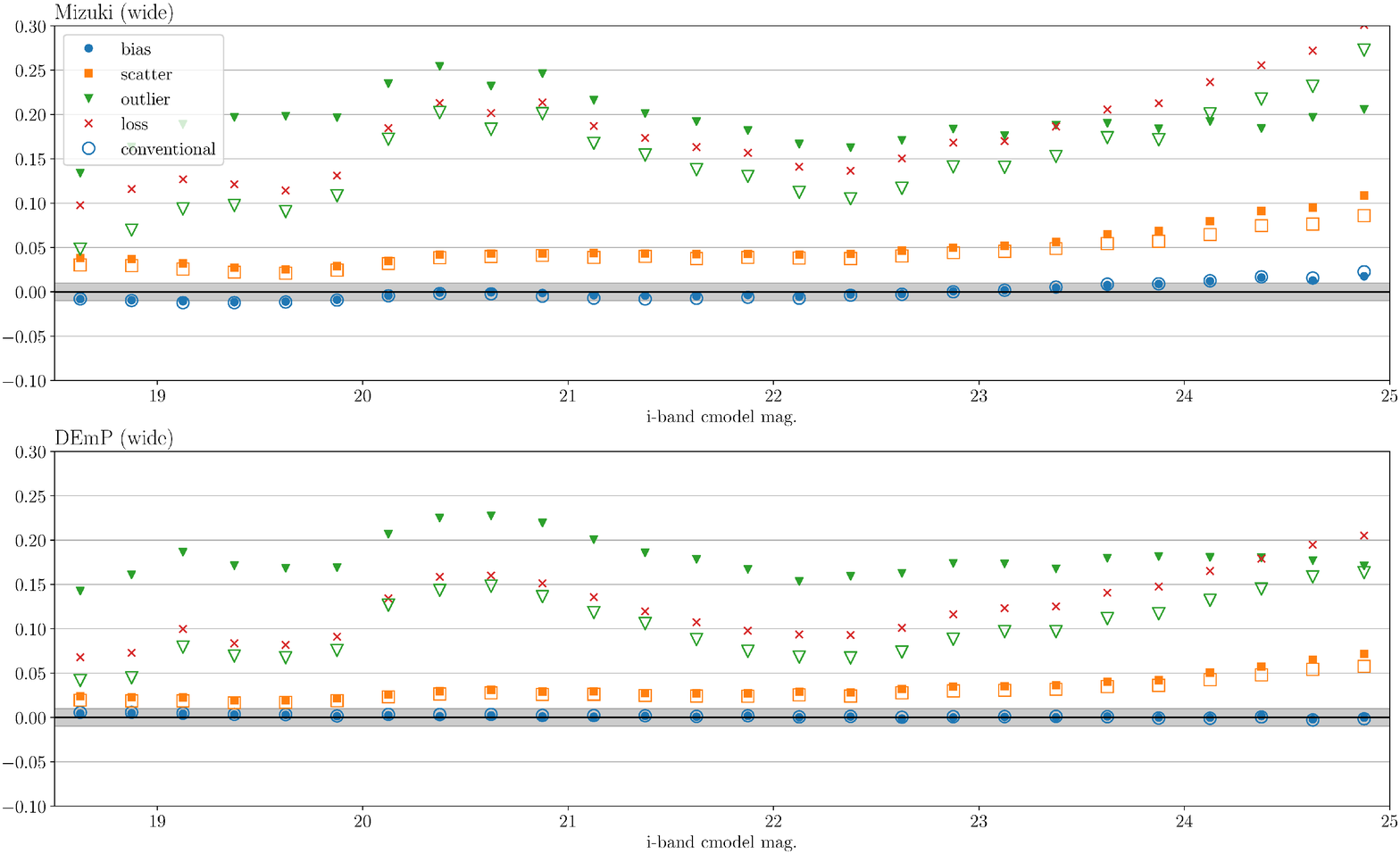}
    \includegraphics[width=\linewidth]{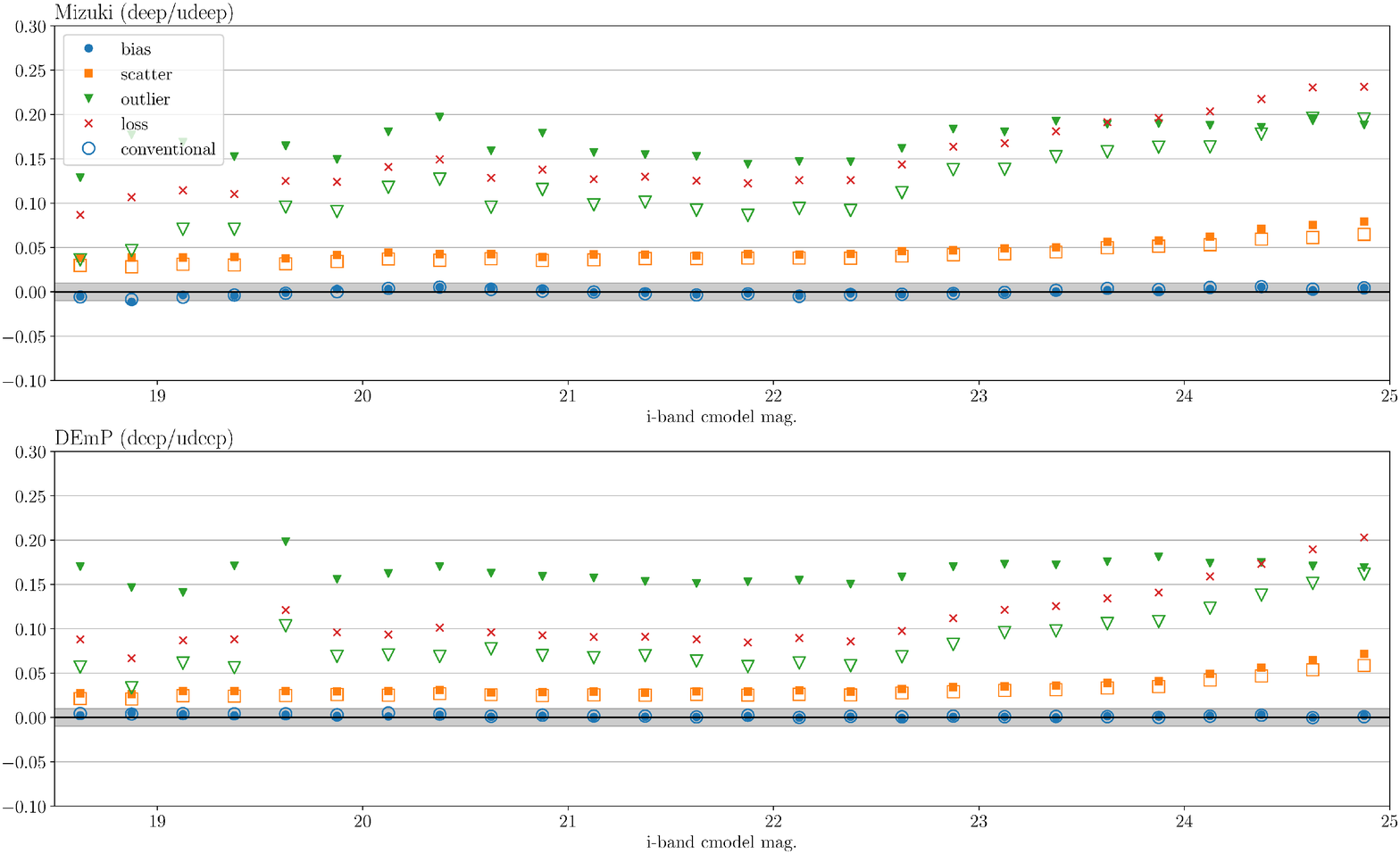}
  \end{center}
  \caption{
    Bias, scatter and outlier rate plotted against $i$-band magnitude.
    The different panels are for different codes as indicated by the label on
    the top-left corner of each panel.  The gray shades show $\pm0.01$ range,
    which will be useful for bias.  The symbols are explained in the panels.
    {\bf
      Note that these plots are based on the COSMOS Wide-depth median stack
      and include objects in COSMOS only.
    }
 }
 \label{fig:stat_mag}
\end{figure*}

\begin{figure*}
  \begin{center}
    \includegraphics[width=\linewidth]{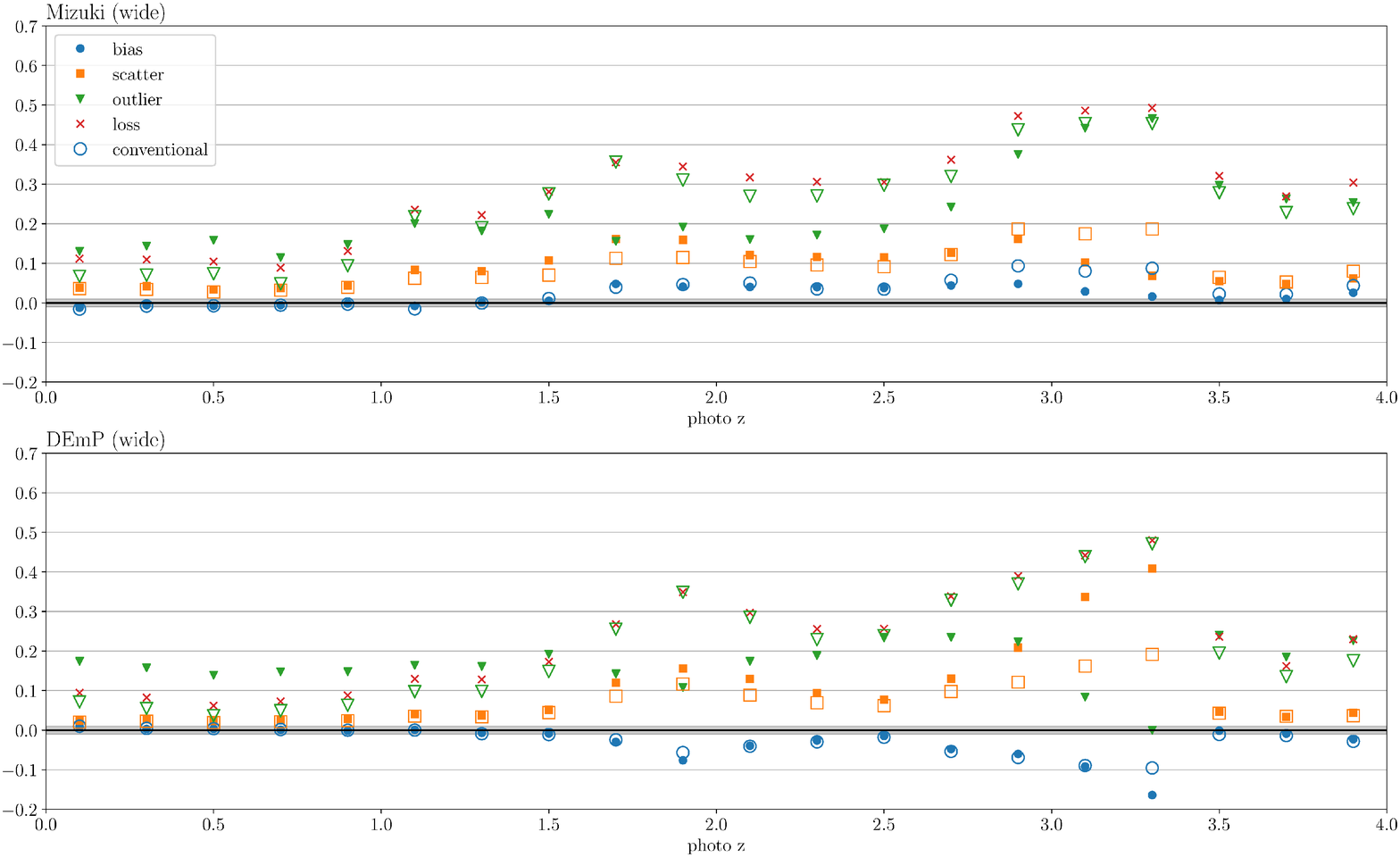}
    \includegraphics[width=\linewidth]{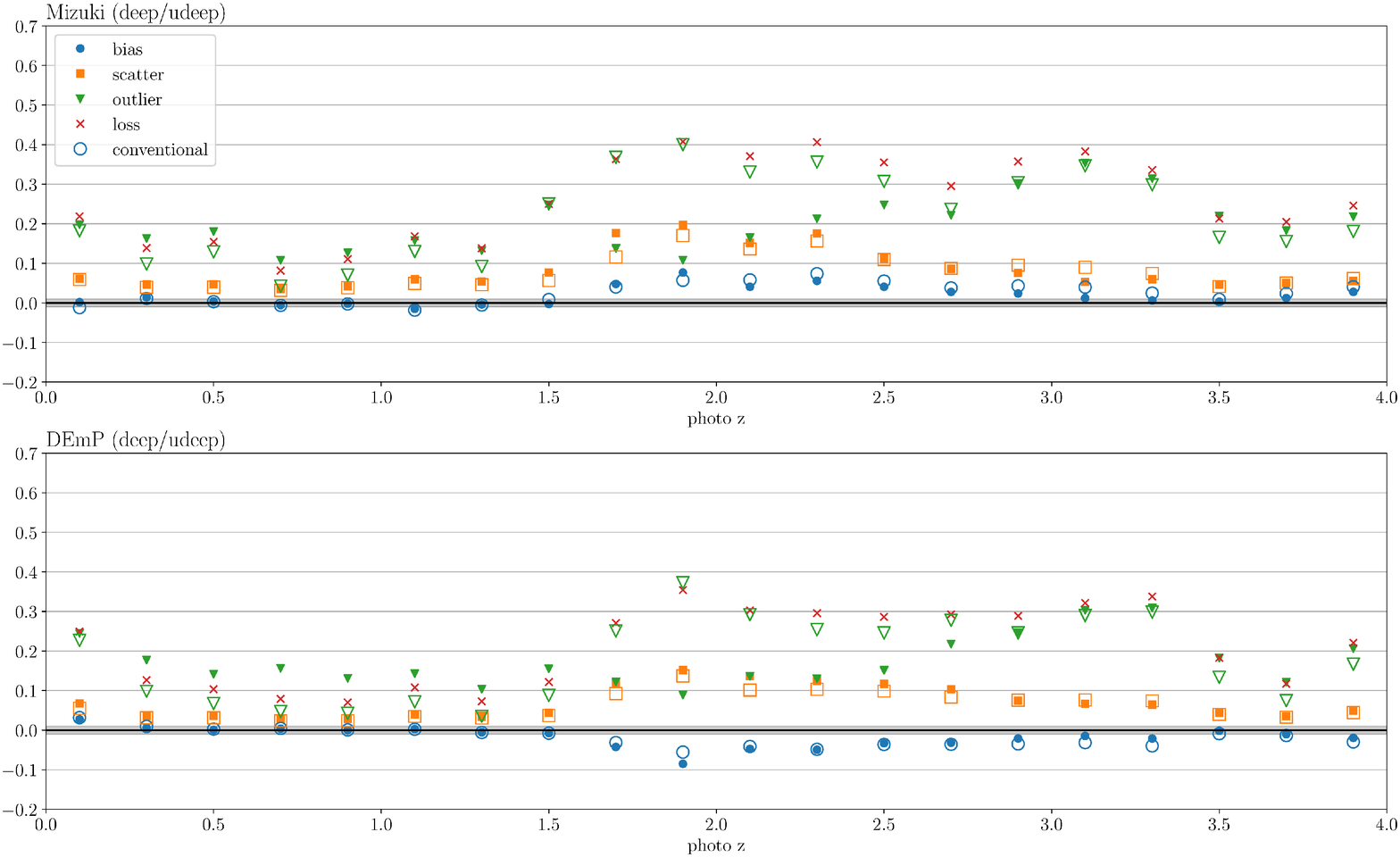}
  \end{center}
  \caption{
    Same as Fig.~\ref{fig:stat_mag} but as a function of \texttt{photoz\_best}.
 }
 \label{fig:stat_zphot}
\end{figure*}

{\scriptsize
  \begin{longtable}{ll|cccc|cccc}
    \caption{
      Photo-$z$ statistics for all the codes as a function of magnitude.
    }
    \label{tab:stat_mag}
    \hline
           &        &  \multicolumn{4}{c|}{wide} & \multicolumn{4}{c}{deep/u-deep} \\
    Code   &  mag.  &  bias  &  $\sigma_{\rm bw}$ & $f_{\rm bw}$ & $<L(\Delta z))>$  &  bias  &  $\sigma_{\rm bw}$ & $f_{\rm bw}$ & $<L(\Delta z))>$ \\
    \endfirsthead
    \endhead
    \hline
    \endfoot
    \hline
    \endlastfoot
    \hline
     & $18.50-18.75$ & $-0.008$  & 0.038  & 0.134  & 0.098  & $-0.005$  & 0.038  & 0.129  & 0.087 \\
 & $18.75-19.00$ & $-0.009$  & 0.037  & 0.163  & 0.116  & $-0.011$  & 0.039  & 0.177  & 0.107 \\
 & $19.00-19.25$ & $-0.011$  & 0.032  & 0.189  & 0.127  & $-0.004$  & 0.039  & 0.169  & 0.115 \\
 & $19.25-19.50$ & $-0.012$  & 0.027  & 0.197  & 0.121  & $-0.004$  & 0.039  & 0.153  & 0.110 \\
 & $19.50-19.75$ & $-0.011$  & 0.025  & 0.198  & 0.114  & $-0.001$  & 0.038  & 0.165  & 0.125 \\
 & $19.75-20.00$ & $-0.009$  & 0.029  & 0.197  & 0.131  & $+0.003$  & 0.042  & 0.149  & 0.124 \\
 & $20.00-20.25$ & $-0.004$  & 0.035  & 0.235  & 0.185  & $+0.003$  & 0.044  & 0.181  & 0.141 \\
 & $20.25-20.50$ & $-0.001$  & 0.042  & 0.254  & 0.213  & $+0.005$  & 0.043  & 0.197  & 0.149 \\
 & $20.50-20.75$ & $-0.001$  & 0.043  & 0.232  & 0.202  & $+0.004$  & 0.043  & 0.159  & 0.129 \\
 & $20.75-21.00$ & $-0.001$  & 0.043  & 0.246  & 0.213  & $+0.002$  & 0.039  & 0.179  & 0.138 \\
 & $21.00-21.25$ & $-0.004$  & 0.044  & 0.216  & 0.187  & $-0.001$  & 0.042  & 0.157  & 0.127 \\
 & $21.25-21.50$ & $-0.004$  & 0.043  & 0.201  & 0.174  & $-0.001$  & 0.042  & 0.155  & 0.130 \\
 Mizuki & $21.50-21.75$ & $-0.005$  & 0.043  & 0.192  & 0.163  & $-0.003$  & 0.041  & 0.153  & 0.125 \\
 & $21.75-22.00$ & $-0.003$  & 0.043  & 0.182  & 0.157  & $-0.002$  & 0.043  & 0.144  & 0.122 \\
 & $22.00-22.25$ & $-0.005$  & 0.042  & 0.167  & 0.141  & $-0.004$  & 0.042  & 0.147  & 0.126 \\
 & $22.25-22.50$ & $-0.003$  & 0.043  & 0.163  & 0.136  & $-0.002$  & 0.043  & 0.147  & 0.126 \\
 & $22.50-22.75$ & $-0.002$  & 0.046  & 0.171  & 0.150  & $-0.003$  & 0.046  & 0.162  & 0.144 \\
 & $22.75-23.00$ & $+0.001$  & 0.050  & 0.184  & 0.168  & $-0.001$  & 0.047  & 0.184  & 0.164 \\
 & $23.00-23.25$ & $+0.002$  & 0.052  & 0.176  & 0.170  & $-0.001$  & 0.049  & 0.181  & 0.168 \\
 & $23.25-23.50$ & $+0.005$  & 0.056  & 0.188  & 0.186  & $+0.000$  & 0.050  & 0.193  & 0.181 \\
 & $23.50-23.75$ & $+0.007$  & 0.065  & 0.190  & 0.206  & $+0.002$  & 0.057  & 0.189  & 0.191 \\
 & $23.75-24.00$ & $+0.009$  & 0.069  & 0.184  & 0.213  & $+0.001$  & 0.058  & 0.190  & 0.196 \\
 & $24.00-24.25$ & $+0.012$  & 0.080  & 0.192  & 0.236  & $+0.003$  & 0.062  & 0.188  & 0.203 \\
 & $24.25-24.50$ & $+0.016$  & 0.091  & 0.184  & 0.255  & $+0.005$  & 0.071  & 0.186  & 0.218 \\
 & $24.50-24.75$ & $+0.013$  & 0.095  & 0.197  & 0.272  & $+0.002$  & 0.075  & 0.196  & 0.231 \\
 & $24.75-25.00$ & $+0.018$  & 0.109  & 0.206  & 0.301  & $+0.004$  & 0.079  & 0.188  & 0.231 \\
\hline
 & $18.50-18.75$ & $+0.004$  & 0.024  & 0.143  & 0.068  & $+0.002$  & 0.027  & 0.170  & 0.088 \\
 & $18.75-19.00$ & $+0.005$  & 0.023  & 0.161  & 0.073  & $+0.005$  & 0.026  & 0.146  & 0.067 \\
 & $19.00-19.25$ & $+0.004$  & 0.023  & 0.186  & 0.100  & $+0.003$  & 0.030  & 0.141  & 0.087 \\
 & $19.25-19.50$ & $+0.003$  & 0.019  & 0.171  & 0.084  & $+0.002$  & 0.030  & 0.171  & 0.088 \\
 & $19.50-19.75$ & $+0.003$  & 0.020  & 0.168  & 0.082  & $+0.003$  & 0.030  & 0.198  & 0.121 \\
 & $19.75-20.00$ & $+0.001$  & 0.022  & 0.169  & 0.091  & $+0.002$  & 0.029  & 0.156  & 0.096 \\
 & $20.00-20.25$ & $+0.002$  & 0.026  & 0.207  & 0.134  & $+0.001$  & 0.030  & 0.162  & 0.093 \\
 & $20.25-20.50$ & $+0.001$  & 0.030  & 0.225  & 0.159  & $+0.003$  & 0.031  & 0.170  & 0.101 \\
 & $20.50-20.75$ & $+0.002$  & 0.031  & 0.227  & 0.160  & $+0.001$  & 0.028  & 0.163  & 0.096 \\
 & $20.75-21.00$ & $+0.001$  & 0.029  & 0.219  & 0.151  & $+0.002$  & 0.028  & 0.159  & 0.093 \\
 & $21.00-21.25$ & $+0.001$  & 0.030  & 0.201  & 0.136  & $+0.000$  & 0.029  & 0.157  & 0.091 \\
 & $21.25-21.50$ & $+0.002$  & 0.027  & 0.186  & 0.120  & $+0.001$  & 0.028  & 0.153  & 0.091 \\
 DEmP & $21.50-21.75$ & $+0.001$  & 0.027  & 0.178  & 0.107  & $+0.001$  & 0.029  & 0.151  & 0.088 \\
 & $21.75-22.00$ & $+0.002$  & 0.027  & 0.167  & 0.098  & $+0.001$  & 0.029  & 0.153  & 0.085 \\
 & $22.00-22.25$ & $+0.000$  & 0.029  & 0.153  & 0.094  & $-0.000$  & 0.030  & 0.155  & 0.090 \\
 & $22.25-22.50$ & $+0.001$  & 0.028  & 0.159  & 0.093  & $+0.001$  & 0.029  & 0.150  & 0.086 \\
 & $22.50-22.75$ & $-0.001$  & 0.032  & 0.163  & 0.101  & $-0.001$  & 0.032  & 0.158  & 0.097 \\
 & $22.75-23.00$ & $-0.000$  & 0.035  & 0.174  & 0.116  & $+0.001$  & 0.034  & 0.170  & 0.112 \\
 & $23.00-23.25$ & $+0.001$  & 0.035  & 0.173  & 0.123  & $+0.001$  & 0.035  & 0.173  & 0.121 \\
 & $23.25-23.50$ & $+0.000$  & 0.036  & 0.168  & 0.125  & $-0.000$  & 0.036  & 0.172  & 0.126 \\
 & $23.50-23.75$ & $+0.001$  & 0.040  & 0.180  & 0.141  & $+0.001$  & 0.039  & 0.176  & 0.134 \\
 & $23.75-24.00$ & $-0.000$  & 0.042  & 0.182  & 0.148  & $+0.001$  & 0.041  & 0.181  & 0.141 \\
 & $24.00-24.25$ & $-0.000$  & 0.050  & 0.180  & 0.165  & $+0.002$  & 0.049  & 0.174  & 0.159 \\
 & $24.25-24.50$ & $+0.002$  & 0.057  & 0.180  & 0.179  & $+0.003$  & 0.056  & 0.175  & 0.173 \\
 & $24.50-24.75$ & $-0.002$  & 0.065  & 0.177  & 0.195  & $-0.000$  & 0.065  & 0.171  & 0.190 \\
 & $24.75-25.00$ & $-0.000$  & 0.072  & 0.171  & 0.205  & $+0.002$  & 0.071  & 0.169  & 0.203 \\
\hline
    \hline
  \end{longtable}
}

{\scriptsize
  \begin{longtable}{ll|cccc|cccc}
    \caption{
      Photo-$z$ statistics for all the codes as a function of \texttt{photoz\_best}. All those numbers are for the sample down to $i=25$ mag.
    }
    \label{tab:stat_zp}
       &        &  \multicolumn{4}{c|}{wide} & \multicolumn{4}{c}{deep/u-deep} \\
Code   &  $z_{\rm phot}$  &  bias  &  $\sigma_{\rm bw}$ & $f_{\rm bw}$ & $<L(\Delta z))>$  &  bias  &  $\sigma_{\rm bw}$ & $f_{\rm bw}$ & $<L(\Delta z))>$ \\
    \endfirsthead
    \endhead
    \hline
    \endfoot
    \hline
    \endlastfoot
    \hline
     & $0.00-0.20$ & $-0.012$  & 0.039  & 0.131  & 0.112  & $+0.002$  & 0.061  & 0.198  & 0.219 \\
 & $0.20-0.40$ & $-0.006$  & 0.042  & 0.144  & 0.110  & $+0.014$  & 0.046  & 0.163  & 0.139 \\
 & $0.40-0.60$ & $-0.008$  & 0.034  & 0.159  & 0.104  & $+0.004$  & 0.047  & 0.180  & 0.154 \\
 & $0.60-0.80$ & $-0.006$  & 0.037  & 0.115  & 0.089  & $-0.006$  & 0.035  & 0.108  & 0.081 \\
 & $0.80-1.00$ & $-0.001$  & 0.044  & 0.148  & 0.131  & $-0.002$  & 0.042  & 0.127  & 0.111 \\
 & $1.00-1.20$ & $-0.008$  & 0.083  & 0.201  & 0.235  & $-0.015$  & 0.059  & 0.158  & 0.168 \\
 & $1.20-1.40$ & $+0.002$  & 0.080  & 0.182  & 0.222  & $-0.004$  & 0.054  & 0.132  & 0.138 \\
 & $1.40-1.60$ & $+0.005$  & 0.107  & 0.224  & 0.281  & $-0.003$  & 0.076  & 0.248  & 0.250 \\
 & $1.60-1.80$ & $+0.048$  & 0.162  & 0.156  & 0.355  & $+0.048$  & 0.176  & 0.138  & 0.364 \\
Mizuki & $1.80-2.00$ & $+0.041$  & 0.159  & 0.192  & 0.344  & $+0.077$  & 0.197  & 0.108  & 0.407 \\
 & $2.00-2.20$ & $+0.040$  & 0.121  & 0.161  & 0.317  & $+0.041$  & 0.152  & 0.165  & 0.371 \\
 & $2.20-2.40$ & $+0.040$  & 0.116  & 0.172  & 0.306  & $+0.056$  & 0.176  & 0.213  & 0.406 \\
 & $2.40-2.60$ & $+0.038$  & 0.115  & 0.188  & 0.305  & $+0.041$  & 0.112  & 0.248  & 0.355 \\
 & $2.60-2.80$ & $+0.044$  & 0.127  & 0.242  & 0.362  & $+0.028$  & 0.086  & 0.221  & 0.295 \\
 & $2.80-3.00$ & $+0.048$  & 0.161  & 0.376  & 0.472  & $+0.024$  & 0.076  & 0.302  & 0.357 \\
 & $3.00-3.20$ & $+0.029$  & 0.102  & 0.442  & 0.486  & $+0.012$  & 0.053  & 0.353  & 0.383 \\
 & $3.20-3.40$ & $+0.016$  & 0.069  & 0.466  & 0.493  & $+0.006$  & 0.060  & 0.315  & 0.336 \\
 & $3.40-3.60$ & $+0.007$  & 0.054  & 0.297  & 0.321  & $+0.003$  & 0.046  & 0.219  & 0.213 \\
 & $3.60-3.80$ & $+0.010$  & 0.048  & 0.263  & 0.269  & $+0.012$  & 0.050  & 0.183  & 0.205 \\
 & $3.80-4.00$ & $+0.026$  & 0.062  & 0.254  & 0.304  & $+0.028$  & 0.055  & 0.218  & 0.246 \\
\hline
 & $0.00-0.20$ & $+0.009$  & 0.024  & 0.174  & 0.094  & $+0.028$  & 0.067  & 0.245  & 0.249 \\
 & $0.20-0.40$ & $+0.003$  & 0.027  & 0.158  & 0.082  & $+0.006$  & 0.036  & 0.177  & 0.127 \\
 & $0.40-0.60$ & $+0.003$  & 0.023  & 0.139  & 0.062  & $+0.001$  & 0.036  & 0.141  & 0.104 \\
 & $0.60-0.80$ & $+0.001$  & 0.024  & 0.148  & 0.072  & $+0.004$  & 0.028  & 0.156  & 0.079 \\
 & $0.80-1.00$ & $-0.001$  & 0.028  & 0.148  & 0.088  & $+0.000$  & 0.028  & 0.131  & 0.070 \\
 & $1.00-1.20$ & $+0.001$  & 0.041  & 0.164  & 0.130  & $+0.003$  & 0.040  & 0.143  & 0.108 \\
 & $1.20-1.40$ & $-0.007$  & 0.037  & 0.161  & 0.128  & $-0.005$  & 0.034  & 0.104  & 0.073 \\
 & $1.40-1.60$ & $-0.008$  & 0.051  & 0.193  & 0.173  & $-0.008$  & 0.044  & 0.155  & 0.122 \\
 & $1.60-1.80$ & $-0.030$  & 0.120  & 0.142  & 0.268  & $-0.043$  & 0.120  & 0.122  & 0.271 \\
DEmP & $1.80-2.00$ & $-0.076$  & 0.155  & 0.108  & 0.349  & $-0.085$  & 0.152  & 0.089  & 0.354 \\
 & $2.00-2.20$ & $-0.039$  & 0.129  & 0.174  & 0.297  & $-0.048$  & 0.136  & 0.136  & 0.303 \\
 & $2.20-2.40$ & $-0.026$  & 0.094  & 0.189  & 0.256  & $-0.050$  & 0.124  & 0.130  & 0.296 \\
 & $2.40-2.60$ & $-0.015$  & 0.077  & 0.235  & 0.256  & $-0.033$  & 0.117  & 0.151  & 0.286 \\
 & $2.60-2.80$ & $-0.048$  & 0.130  & 0.234  & 0.338  & $-0.032$  & 0.103  & 0.218  & 0.293 \\
 & $2.80-3.00$ & $-0.060$  & 0.209  & 0.223  & 0.389  & $-0.021$  & 0.075  & 0.246  & 0.289 \\
 & $3.00-3.20$ & $-0.092$  & 0.337  & 0.084  & 0.442  & $-0.015$  & 0.067  & 0.303  & 0.321 \\
 & $3.20-3.40$ & $-0.164$  & 0.408  & 0.000  & 0.480  & $-0.021$  & 0.065  & 0.309  & 0.338 \\
 & $3.40-3.60$ & $-0.001$  & 0.046  & 0.240  & 0.236  & $-0.002$  & 0.044  & 0.182  & 0.183 \\
 & $3.60-3.80$ & $-0.009$  & 0.033  & 0.185  & 0.162  & $-0.010$  & 0.035  & 0.122  & 0.117 \\
 & $3.80-4.00$ & $-0.023$  & 0.044  & 0.226  & 0.230  & $-0.020$  & 0.049  & 0.206  & 0.221 \\
\hline
    \hline
  \end{longtable}
}

\subsection{Code-code comparisons for Point Estimate}
\label{ssec:code_code_comparisons}
In this section, we present the \photoz\, performance for \texttt{Mizuki} and \texttt{DEmP} based on the point statistics by using the $k=5$ validation set. We show both Wide and Deep/UltraDeep performances and for the sample matched only to the HSC Deep/UltlaDeep, the \photoz\, is derived according to the wide-emulated photometric errors.

Figure \ref{fig:scatterPlot} shows the scatter plot of PDR2 \photoz\, for two different codes for two different depths. We note that for the wide depth, we use not only the objects detected in the wide layer but also objects detected in Deep/UltraDeep layers as well replacing the photometric error with the wide-emulated errors when observed wide photometric error is not available. If we compare the Wide and Deep/UltraDeep depths, we can see little difference between them. This implies that the accuracy of photo-z is already converged at the Wide depth of the HSC and going deeper observation would little help to improve the accuracy. The statistics defined in section \ref{ssec:metric} are shown in Figures \ref{fig:stat_mag} and \ref{fig:stat_zphot}.
Top and 3rd panels of Fig. \ref{fig:stat_mag} show the statistics for \texttt{Mizuki} for Wide and Deep/UltraDeep catalogs. Outlier rate for all magnitude ranges and the bias at faint magnitudes $i\sim 25$ are slightly improved with deeper photometry. Therefore, the deeper photometry is still useful for the photo-z at faint objects, $i>24$.
The second and bottom panels show the statistics for \texttt{DEmP}. No significant improvements are observed in bias and scatter, but the outlier rate is slightly improved at all magnitude ranges.
Figure \ref{fig:stat_zphot} shows the statistics as a function of \texttt{photoz\_best}. For both codes, the deeper image help to improve at redshift range $2<z_p<3.5$.

\subsection{N(z) distribution}
\label{ssec:dndz}

In various scientific uses, we often consider not only the redshift
for single galaxy but also the global properties averaged over a
number of objects. In this section, we show redshift distributions of
photometric sample and compare them between two different photo-$z$ codes.

\begin{figure}
  \includegraphics[width=\linewidth]{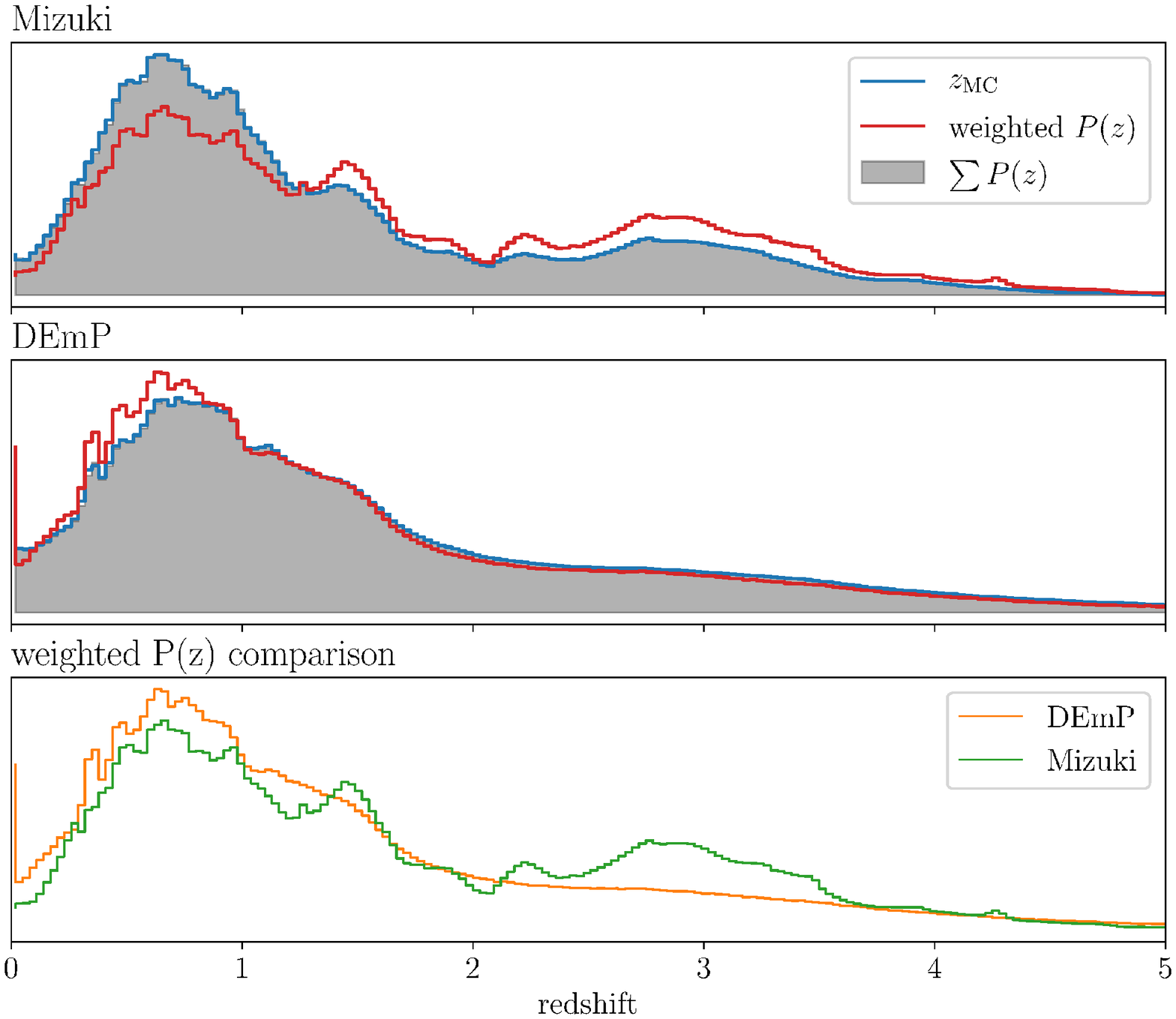}
  \caption{Redshift distributions for entire PDR2 wide sample computed from $z_{\rm MC}$ (blue) and stacked $P(z)$ (gray shaded). Also shown with red histogram is reweighted distribution of the reference redshift sample. Bottom panel show the comparison of the reweighted distribution for \texttt{Mizuki} and \texttt{DEmP}.
    \label{fig:dndz_wide}}
\end{figure}

Figure \ref{fig:dndz_wide} shows the $N(z)$ distribution for the entire PDR2 catalog in the wide layer. First we compare the distributions by stacking the $P(z)$ for all galaxies and the histogram of $z_{\rm MC}$ a random draw from the $P(z)$. As clarified in the PDR1 paper, these two different measurements fairly well agree with each other. 
Also shown with red curves is the one obtained by reference redshift catalog with reweighting to match the magnitude and multi-color distribution to the one for the entire photometric sample in the wide layer. For \texttt{DEmP}, the weighted distribution agrees well with the stacked $P(z)$, while the significant disagreement can be seen for the \texttt{Mizuki} catalog. This is mainly due to the fact that the \texttt{Mizuki} only have photo-z for objects observed with three bands or more (see Table \ref{tab:photoz_depend}). Therefore, the weights to apply to the reference redshift catalog do not reproduce the entire photometric sample of \texttt{Mizuki}. 
We note that this disagreement does not mean that \texttt{Mizuki} worse measure the photo-$z$ but simply means that the mismatch between the target weight and the actual \texttt{Mizuki}'s sample.
The bottom panel compares the difference in weighted $P(z)$ between \texttt{Mizuki} and \texttt{DEmP}. Because of the stringent sample selection of \texttt{Mizuki}, the distribution is slightly shallower compared to the one from \texttt{DEmP}; mean redshifts are 1.44 and 1.68 and medians 1.1 and 1.37 for \texttt{Mizuki} and \texttt{DEmP} respectively.

\subsection{Tests on PDF}
\label{ssec:test_on_pdf}
The full information of photo-z is encoded in the probability distribution function (PDF), or in other words, the posterior distribution of the measurement for individual galaxy. In this section, we assess the quality of the PDF by exploring with two quantities as in the PDR1 paper. The statistics we focus on are probability integral transform (PIT) and continuous ranked probability score (CRPS) \citep{polsterer16}.

\begin{figure*}
  \begin{center}
    \includegraphics[width=\linewidth]{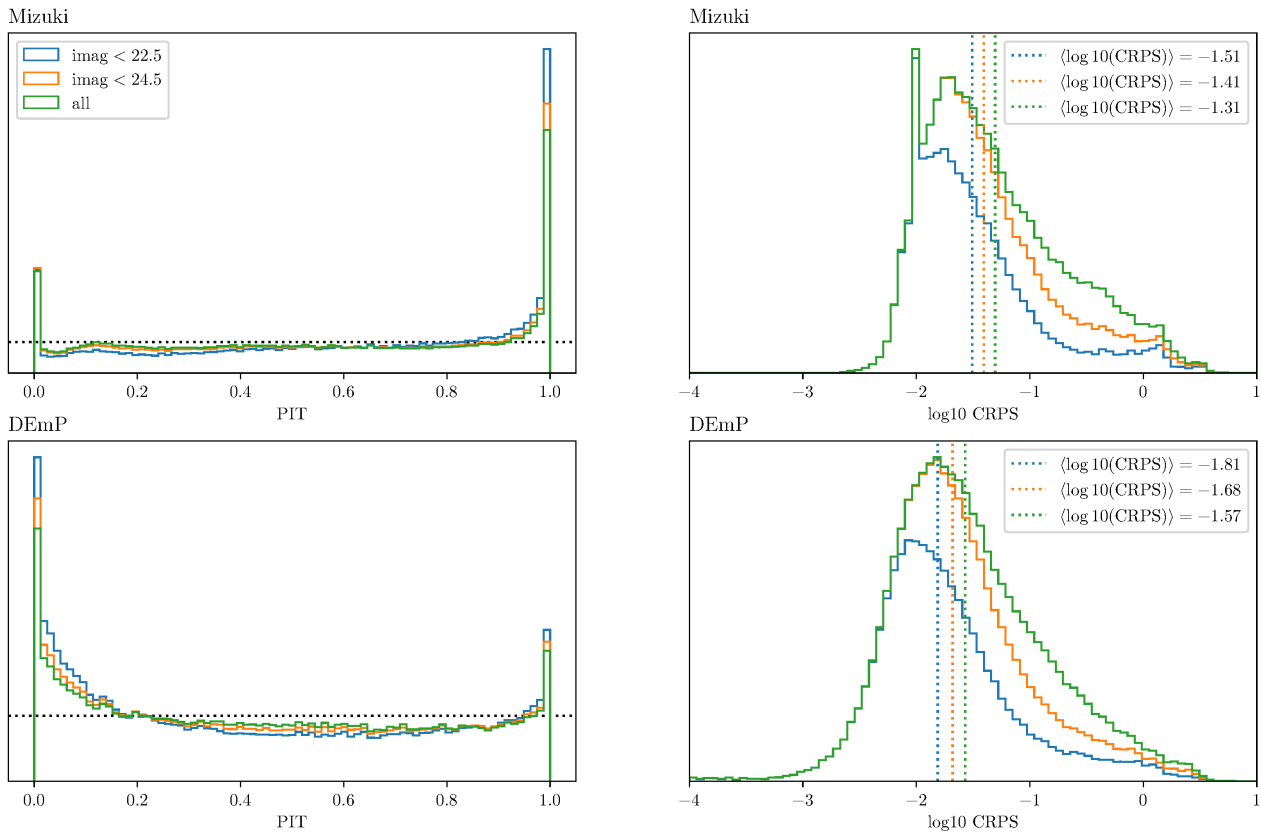}
  \end{center}
  \caption{
    PIT (left) and CRPS (right) for \texttt{Mizuki} (upper) and \texttt{DEmP} (lower).  The dotted horizontal line in the left panel is just to guide the eye, and dotted lines in the right panels show the mean of CRPS.
 }
 \label{fig:test_on_pdf}
\end{figure*}

The PIT analysis evaluates if the PDF correctly reflects the error of the photo-z. The PIT is defined as 
\begin{equation}
    {\rm PIT}(z_{\rm ref}) 
    = 
    \int_0^{z_{\rm ref}} P(z) dz,
\end{equation}
and we take the histograrm of PIT over the test sample.
When the PDF is properly calibrated, the histogram will become flat distribution. The left panels of Fig. \ref{fig:test_on_pdf} represent the PIT distributions normalized to unity for \texttt{Mizuki} (upper) and \texttt{DEmP} (lower). Two significant peaks at PIT$\sim 0$ and $1$ can be observed which indicate that the PDF is too narrow. Note that the \texttt{Mizuki} shows more flatter shape than \texttt{DEmP} because the PDF of \texttt{Mizuki} is recalibrated with the PIT of the training sample \citep{bordoloi10}. In principle this recalibration works fairly well; however in practical case where an extreme situation such that the PDF only has non-zero value at $z=0$, the recalibration does not perfectly make the distribution of PIT flat, and peaks still remain. The recalibration has not been applied for \texttt{DEmP} and this makes PIT distribution more convex shape. 

The CRPS measures the distance between $z_{\rm ref}$ and the PDF, which is defined as
\begin{equation}
    {\rm CRPS} = \int_0^\infty [{\rm PIT}(z) - H(z-z_{\rm ref})]^2 dz,
\end{equation}
where $H(z)$ is the Heaviside step function. The right panels of Fig. \ref{fig:test_on_pdf} show the distribution of CRPS for entire test sample and means of CRPS. In terms of the CRPS, brighter objects ($i<22.5$) have better statistics for both \texttt{Mizuki} and \texttt{DEmP}, and \texttt{DEmP} is slightly better than \texttt{Mizuki}. One can see the strinking peak at $\log10({\rm CRPS})=-2$ for \texttt{Mizuki}, which is dominated by the objects at $z=0$ and PDF only has a single non-zero value at z=0.

\subsection{Comparison to PDR1}
\label{ssec:comp_dr1_dr2}
In this section, we compare the accuracy of the photo-z between PDR1 and PDR2. A major difference between PDR1 and PDR2 is in a background subtraction, but we briefly revisit the difference in the photometric data between two releases.
\begin{itemize}
    \item{} The sky coverage of wide layer full-depth-full-color (FDFC) region increase from 100 to 300 square degrees and we include not only FDFC region, but also the area where the some fraction of color or depth are missing.
    \item{} We do not discriminate Deep and UltraDeep layers and they are jointly processed.
    \item{} The broadband filters $r$ and $i$ are replaced with more uniform transmission ones, $r2$ and $i2$.
    \item{} A new sky subtraction is implemented which may affect the tail of the large objects, and possibly the photometory for objects located in the vicinity of the large object.
    \item{} Object detection algorithm is improved and much fainter objects are detected.
    \item{} Photometric calibration using Pan-STARRS has been updated, and we remove the late-type stars from calibration to avoid the metalicity variations.
\end{itemize}
There are more updates from PDR1 for the photometric data but refer to the second data release paper \citep{aihara19} for more complete statement.

\begin{figure*}
  \begin{center}
    \includegraphics[width=\linewidth]{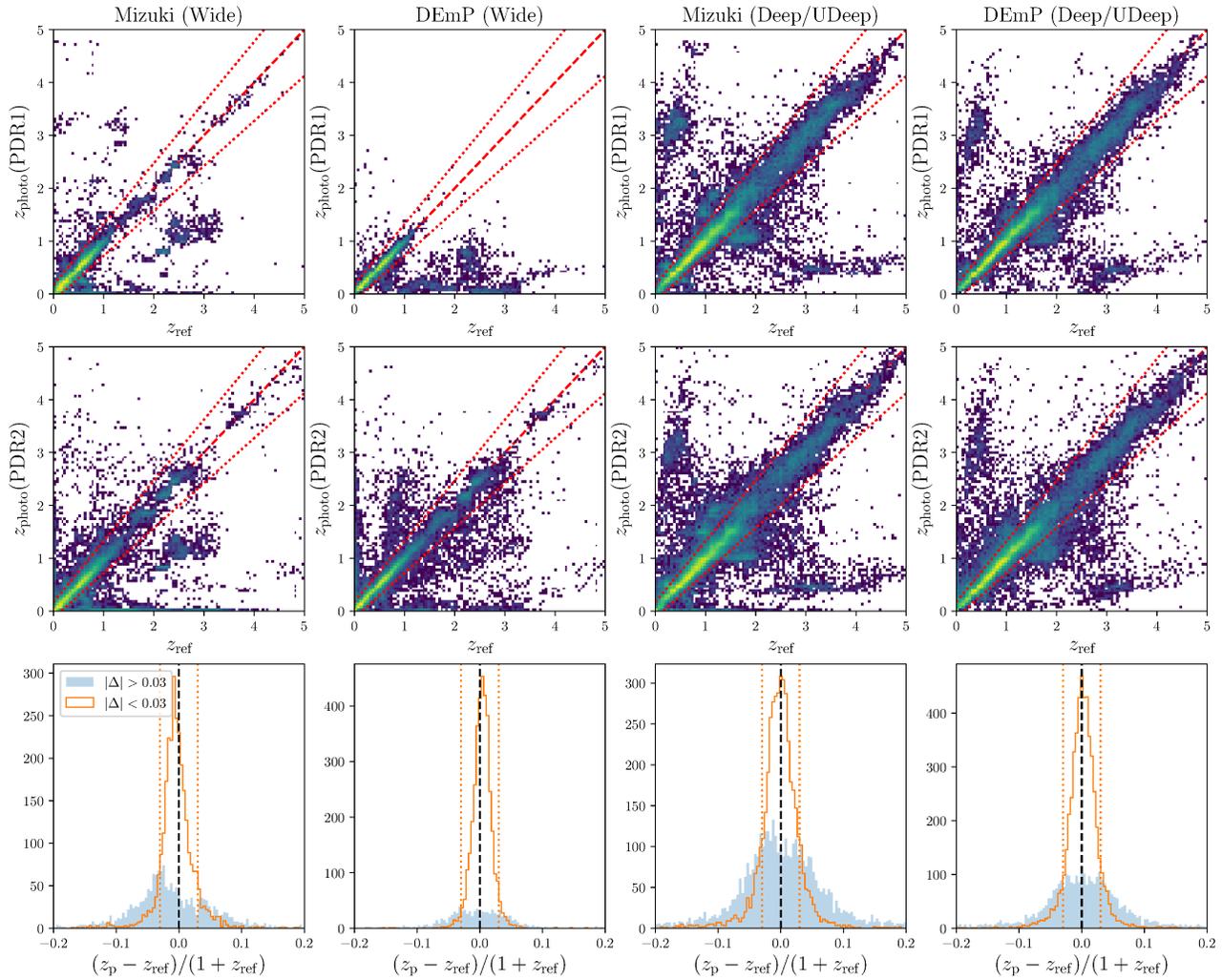}
  \end{center}
  \caption{
    Comparison of photoz accuracy between PDR1 and PDR2. Upper panels are for PDR1 and middle panels for PDR2. A significant extension to the higher redshifts for PDR2 can be observed for the wide layer. The bottom panels show the one to one comparison. Orange open histogram show the accuracy of PDR2 objects that have accuracy better than $\Delta z<0.03$ for PDR1. Blue filled histogram shows the one for objects having worse than 0.03 for PDR1.
 }
 \label{fig:comp_dr1_dr2}
\end{figure*}

Figure \ref{fig:comp_dr1_dr2} shows the difference of photo-z accuracy between PDR1 and PDR2. Upper panels show the scatter plots for the PDR1 test sample, and the middle panels for the PDR2. It is clearly seen that the PDR2 wide sample is extended to the higher redshifts due to the increase of the reference redshift sample. There are no remarkable improvement for the Deep and UltraDeep objects.

Next, we match the objects in PDR2 to those in PDR1 by coordinate. 4k objects out of 26k objects in the test sample are matched within 1 arcsec for wide, and 9k objects out of 49k objects for Deep/UltraDeep. This low probability of object matching is simply because we generate the test sample for PDR2 randomly and independently from PDR1: roughly speaking, 1/5 objects should be matched. The bottom panels of Fig. \ref{fig:comp_dr1_dr2} show the photo-z accuracy of matched PDR2 sample for objects having better accuracy at PDR1 ($\Delta z<0.03$) and worse accuracy ($\Delta z > 0.03$) respectively. The orange open histogram show tails extended to $\Delta z>0.03$ which means some fraction of objects for better photo-z at PDR1 gets worse at PDR2. On the other hand, objects having worse accuracy at PDR1 (shown in filled blue histogram) has a peak at $\Delta z=0$, which reflects that the large fraction of objects are getting better accuracy in PDR2. For \texttt{Mizuki}, 45\% (42\%) objects are improved and 15\% (7\%) objects remain same accuracy for Wide (Deep/UltraDeep) sample. For \texttt{DEmP}, 36\% (40\%) objects are improved and 21\% (11\%) objects remain same accuracy, respectively.  This implies that the PDR2 does not significantly improve the accuracy of the photo-z and the accuracy is fluctuating depending on the measurement of the photometry and the way of calibration.

\section{Data Products}
\label{sec:products}
This section summarizes our target selection criteria, 'common' outputs that are available for all the codes, as well as code-specific outputs.

HSC-SSP Public Data Release 2 (PDR2) includes our photo-$z$'s for the Wide and Deep/UltraDeep layers, covering
over 1100 and 30 square degrees respectively.
As in the PDR1, we apply different cuts to select objects for computing photo-$z$ production, depending on the specification of the code. Table \ref{tab:photoz_depend} summarizes the target selection by each code.
The table also indicates whether there are additional outputs from the code,
which we will elaborate below. Both codes have common selection of \texttt{detect\_is\_primary}
to select primary objects. \texttt{DEmP} computes photo-$z$'s for all the primary
objects, but \texttt{Mizuki} for primary objects with good CModel
photometry in at least 3 bands.

Both codes first generate a PDF for each object and then  we run a common script to
compute, as a post process, various point estimates, confidence intervals and other
statistics. The common outputs are summarized in Table \ref{tab:photoz_common}.
In addition to these common outputs, there are code-specific outputs as follows.

\noindent
\textbf{Mizuki}
\begin{itemize}
  \item{\texttt{reduced\_chisq}, $\chi^2_\nu$:} Reduced chi-squares of the best-fit model.
    It is recommended to remove objects having $\chi^2_\nu>5$ for
    scientific use.
  \item{\texttt{stellar\_mass}:} Median stellar mass derived from $P(M*)$, which is
    stellar mass PDF marginalized over all the other parameters.
    The 68\% confidence intervals are also available. All the
    uncertainties on physical parameters include uncertainties from
    photo-$z$'s.
  \item{\texttt{sfr}:} Median star formation rate with 68\% intervals.
  \item{\texttt{tauv}, $\tau_V$:} Median dust attenuation in the
    V-band with 68\% intervals. Note that $A_V=1.09\tau_V$.
  \item{\texttt{prob\_x}: } \texttt{x} is either \texttt{gal},
    \texttt{qso} or \texttt{star}, which denote the relative
    probability that an object is galaxy, QSO and star.
  \item{rest-frame magnitudes:} Rest-frame magnitudes in the GALEX,
    SDSS, HSC, and WFCAM filters. Only the magnitudes from the best-fit
    template at the median redshifts are computed and no uncertainties
    are currently available.
\end{itemize}

\noindent
\textbf{DEmP}
\begin{itemize}
    \item{\texttt{stellar\_mass}:} Mode stellar mass derived from $P(M*)$, in the unit of solar mass in log scale.
    \item{\texttt{stellar\_mass\_err95\_min}:} 2.5 \% percentile in the $P(M*)$
    \item{\texttt{stellar\_mass\_err68\_min}:} 16 \% percentile in the $P(M*)$
    \item{\texttt{stellar\_mass\_err68\_max}:} 84 \% percentile in the $P(M*)$
    \item{\texttt{stellar\_mass\_err95\_max}:} 97.5 \% percentile in the $P(M*)$
    \item{\texttt{SFR}:} Mode star formation rate derived from $P(SFR)$, in the unit of solar mass per year in log scale.
    \item{\texttt{SFR\_err95\_min}:} 2.5 \% percentile in the $P(SFR)$
    \item{\texttt{SFR\_err68\_min}:} 16 \% percentile in the $P(SFR)$
    \item{\texttt{SFR\_err68\_max}:} 84 \% percentile in the $P(SFR)$
    \item{\texttt{SFR\_err95\_max}:} 97.5 \% percentile in the $P(SFR)$
\end{itemize}

All of the catalog products such as photo-$z$ point estimates are available in the database.
The full PDFs are stored
in the fits format and are available from the photo-$z$ page of the PDR2 site;
\url{https://hsc-release.mtk.nao.ac.jp/}.

\begin{table*}
  \begin{center}
    \begin{tabular*}{\textwidth}{l @{\extracolsep{\fill}} ll}\hline\hline
      \texttt{key}                & description \\ \hline \hline
      \texttt{object\_id}         & unique object id to be used to join with the photometry tables \\
      \texttt{photoz\_X}          & Photo-$z$ point estimate where X is
                                    either \texttt{mean}, \texttt{mode}, \texttt{median}, or \texttt{best} \\
      \texttt{photoz\_mc}         & Monte Carlo draw from the full PDF \\
      \texttt{photoz\_conf\_X}    & Photo-$z$ confidence value for \texttt{photoz\_X}, defined
                                    by equation (15) of PDR1 paper \\
      \texttt{photoz\_risk\_X}    & Risk parameter for \texttt{photoz\_X}, defined by equation
                                    (13) of PDR1 paper \\
      \texttt{photoz\_std\_X}     &  Second order moment around a point estimate (\texttt{photoz\_X}) derived from full PDF.\\
      \texttt{photoz\_err68\_min} & 16 \% percentile in the PDF\\
      \texttt{photoz\_err68\_max} & 84 \% percentile in the PDF\\
      \texttt{photoz\_err95\_min} & 2.5 \% percentile in the PDF\\
      \texttt{photoz\_err95\_max} & 97.5 \% percentile in the PDF\\ \hline \hline
    \end{tabular*}
  \end{center}
  \caption{
    Common photo-$z$ parameters available for all the codes.\\
    \label{tab:photoz_common}
  }
\end{table*}

\begin{table*}
  \begin{center}
    \begin{tabular*}{\textwidth}{l @{\extracolsep{\fill}} llll}\hline\hline
      \texttt{CODE} &
                 target selection &
                 number of objects &
                 other quantities \\ \hline \hline
      \texttt{DEmP} &
                 \texttt{detect\_is\_primary} is \texttt{True} &
                 451,520,387 (Wide) &
                 None   \\
                 &  & 20,451,226 (Deep/UltraDeep) & \\ \hline
      \texttt{Mizuki} &
                 \texttt{detect\_is\_primary} is \texttt{True} &
                 137,162,491 (Wide) &
                 many \\
                       &
                 objects with CModel fluxes in at least three bands &
                 19,340,872 (Deep/UltraDeep)
                       \\ \hline \hline
    \end{tabular*}
  \end{center}
  \caption{
    Target selection applied by each code. The number of objects
    that satisfy the selection is shown.
    Details of other  quantities available in the catalog can be found
    in Section \ref{sec:products}.
    \label{tab:photoz_depend}
  }
\end{table*}

\section*{Acknowledgment}
The Hyper Suprime-Cam (HSC) collaboration includes the astronomical communities of Japan and Taiwan,
and Princeton University.  The HSC instrumentation and software were developed by the National
Astronomical Observatory of Japan (NAOJ), the Kavli Institute for the Physics and Mathematics of
the Universe (Kavli IPMU), the University of Tokyo, the High Energy Accelerator Research Organization (KEK),
the Academia Sinica Institute for Astronomy and Astrophysics in Taiwan (ASIAA), and Princeton University.
Funding was contributed by the FIRST program from Japanese Cabinet Office, the Ministry of Education,
Culture, Sports, Science and Technology (MEXT), the Japan Society for the Promotion of Science (JSPS),
Japan Science and Technology Agency  (JST),  the Toray Science  Foundation, NAOJ, Kavli IPMU, KEK, ASIAA,
and Princeton University.

This paper makes use of software developed for the Large Synoptic Survey Telescope. We thank the LSST
Project for making their code available as free software at http://dm.lsst.org.

The Pan-STARRS1 Surveys (PS1) have been made possible through contributions of the Institute for Astronomy,
the University of Hawaii, the Pan-STARRS Project Office, the Max-Planck Society and its participating institutes, the Max Planck Institute for Astronomy,
Heidelberg and the Max Planck Institute for Extraterrestrial Physics, Garching, The Johns Hopkins University, Durham University, the University of Edinburgh,
Queen's University Belfast, the Harvard-Smithsonian Center for Astrophysics, the Las Cumbres Observatory Global Telescope Network Incorporated,
the National Central University of Taiwan, the Space Telescope Science Institute, the National Aeronautics and Space Administration under Grant No.
NNX08AR22G issued through the Planetary Science Division of the NASA Science Mission Directorate, the National Science Foundation under Grant No. AST-1238877,
the University of Maryland, and Eotvos Lorand University (ELTE) and the Los Alamos National Laboratory.

This paper is based on data collected at the Subaru Telescope and retrieved from the HSC data archive system,
which is operated by Subaru Telescope and Astronomy Data Center at National Astronomical Observatory of Japan.
We thank the COSMOS team for making their private spectroscopic redshift catalog available for our calibrations.
MT acknowledges supported by JSPS KAKENHI Grant Number JP15K17617. AN is supported in part by MEXT KAKENHI
Grant Number 16H01096. 
Any opinion, findings, and conclusions or recommendations expressed in this material are 
those of the authors(s) and do not necessarily reflect the views of the National Science Foundation.

\bibliographystyle{mn2e}
\bibliography{bibdata}


\end{document}